# Using NLP to analyze whether customer statements comply with their inner belief

*Research Paper*

Fabian Thaler*
Doctoral Student
fabian.thaler@hnu.de
+49 731 9762 1546

Neu-Ulm University of
Applied Sciences
Center for Research on
Service Sciences (CROSS)
Faculty of Information
Management
Wileystr. 1
89231 Neu-Ulm
Germany

Stefan Faußer
Professor
stefan.fausser@hnu.de
+49 731 9762 1517

Neu-Ulm University of
Applied Sciences
Wileystr. 1
89231 Neu-Ulm
Germany

Heiko Gewald
Professor
heiko.gewald@hnu.de
+49 731 9762 1521

Neu-Ulm University of
Applied Sciences
Center for Research on
Service Sciences (CROSS)
Faculty of Information
Management
Wileystr. 1
89231 Neu-Ulm
Germany

* corresponding author

**Acknowledgments**

The authors are grateful for the support of the Technology Transfer Center, Günzburg, Germany.

**Declaration of Conflicting Interests**

The authors declare no potential conflicts of interest with respect to the research, authorship, and/or publication of this article.

**Funding**

The authors hereby disclose receipt of the following financial support for the research, authorship, and/or publication of this article: The authors would like to thank Technology Transfer Center, Günzburg, Germany for funding the data collection for this study.

**Abstract**

Customers' emotions play a vital role in the service industry. The better frontline personnel understand the customer, the better the service they can provide. As human emotions generate certain (unintentional) bodily reactions, such as increase in heart rate, sweating, dilation, blushing and paling, which are measurable, artificial intelligence (AI) technologies can interpret these signals.

Great progress has been made in recent years to automatically detect basic emotions like joy, anger etc. Complex emotions, consisting of multiple interdependent basic emotions, are more difficult to identify. One complex emotion which is of great interest to the service industry is difficult to detect: whether a customer is telling the truth or just a story…

This research presents an AI-method for capturing and sensing emotional data. With an accuracy of around 98%, the best trained model was able to detect whether a participant of a debating challenge was arguing for or against her/his conviction, using speech analysis. The data set was collected in an experimental setting with 40 participants.

The findings are applicable to a wide range of service processes and specifically useful for all customer interactions that take place via telephone. The algorithm presented can be applied in any situation where it is helpful for the agent to know whether a customer is speaking to her/his conviction. This could, for example, lead to a reduction in doubtful insurance claims, or untruthful statements in job interviews. This would not only reduce operational losses for service companies, but also encourage customers to be more truthful.

# 1 Introduction

Artificial intelligence (AI), commonly understood as "machines that exhibit aspects of human intelligence" (Huang and Rust 2018, p.155), has fascinated mankind for centuries. First tangible implementations date back to the 18th century, such as the famous Mechanical Turk automatic chess player, which in fact was not artificial but human intelligence. Significant advances began in the 1950s, focusing on knowledge-based systems and artificial neural networks (Buchanan 2005). Major milestones were reached in 1997 when IBM's Deep Blue program defeated the world chess champion and in 2006, when the cognitive AI machine Watson defeated a human Jeopardy! Champion (Huang et al. 2019). Over the last two decades, AI made astonishing leaps forward and the "rise of the machines" (Hollebeek et al. 2021) nowadays affects many aspects of the daily lives of people around the world.

In the service industry, interactions with AI happen all along the value chain: from frontline interactions to customer relationship management to back-office processing activities (Huang and Rust 2018). The corporate intentions behind the deployment of AI in customer interactions are obvious: to serve their customers better (increase retention), sell more products, and reduce operational costs. An example familiar to many customers is Amazon's recommender system, which not only provides its customers with new ideas for what else to buy – generating massive (yet undisclosed) additional revenue for the company – but also spilled over to other areas of the internet wherever people are looking for information, videos, news, etc. (Smith and Linden 2017). Another less popular example is the use of AI in queue management for call centers. No customer is happy to wait for the next available agent so companies try hard to reduce waiting time. For example, robotic process automation (RPA) helps to eliminate redundant customer and employee effort by making use of all available data, while natural language processing (NLP) enables interaction in natural language instead of simply giving the caller a set of pre-defined choices (the traditional interactive voice response (IVR) approach). But NLP can do much more: Emotion analysis can improve the process by detecting emotions like anger or dissatisfaction in the customer's voice (Sudarsan and Kumar 2019), predicting corresponding customer behavior and providing the handling agent upfront with recommendations on how best to deal with the issue (Ponomareff 2017). Customers in severe distress may even be assigned to faster queues and/or to specifically trained agents to de-escalate the situation. This illustrates the value of incorporating customers' emotions into service delivery processes. Further examples of AI-detected customer emotions include the use as a diagnostic tool for therapists (Gideon et al. 2019; Tokuno et al. 2011), for fraud detection (Verschuere et al. 2006), and, of course, for all aspects of marketing (see e.g. Huang and Rust 2021). The work of Mattila and Enz (2002) and Locke (1996) underlines the crucial role of emotions in service encounters.

Literally every customer-facing role could benefit from this type of information, especially if the interaction takes place remotely, e.g., via telephone where non-verbal communication is limited. But the telephone is not the only communication channel where the customer uses her/his voice. Industry experts assume that today already 50% of all internet searches are voice searches and predict that almost half the world's digital consumers could be engaging with voice or digital assistants in the future (DBS Interactive 2020). Perrin (2020) and Griffin (2021) forecast that nearly 90% of smartphone users use voice assistants by 2023. Therefore, capturing voice interactions to analyze the inherent customer emotions will become easier year by year.

Current research on utilizing AI to detect emotions from speech typically focuses on the traditional set of six emotions: joy, love, surprise, sadness, anger, and fear (Shaver et al. 1987). This leaves out an important emotional facet: the question of *whether the person is speaking according to her/his true conviction.* For many service interactions it is of utmost importance to understand whether the customer really believes what s/he is saying, or if s/he is just pretending…

We conducted a controlled experiment to address this question. Placed in the setting of a debating challenge the voice recordings of 40 participants were captured. In pairs of two, the participants had to either argue for or against a randomly assigned popular topic. Using AI methods, we tested a series of models to determine whether a particular speaker is arguing for or against her/his true conviction. The results are encouraging: the best model achieves around 98% accuracy in identifying both positives (argument represents conviction) and negatives (argument does not represent conviction). The applied methods solely analyze the voice of the speaker, without semantical analysis of the spoken words.

Our findings can be applied to all types of verbal customer interactions in service provision. For example, in sales to confirm customer's (real) purchase intention, in job interviews to determine whether the applicant is telling the truth about herself/himself and really wants the job, or in all types of negotiations to find out whether the counterpart believes what s/he is saying or is bluffing.

The remainder of this paper is structured as follows: First, we provide an overview of the related research and derive our research objectives. Thereafter we describe our research methodology and explain the setting of the experiment and the data gathering process. We then present the process of building and testing the AI models, elaborate on the results, discuss potential applications in the service industry, identify the limitations of our study and provide avenues for further research. The paper closes with its conclusion.

# 2 Related Work and Research Objective

## 2.1 The Concept of Emotions

The term 'emotion' refers to a complex construct that has not yet been defined to widespread satisfaction. As Mulligan and Scherer (2012, p.345) state: "There is no commonly agreed-upon definition of emotion in any of the disciplines that study this phenomenon." On broad terms, one could say that "the term 'emotion' exemplifies [an] 'umbrella' concept that includes affective, cognitive, behavioral, expressive and physiological changes" (Tyng et al. 2017, p.3). Thus, "Emotion is a complex state that combines feelings, thoughts, and behavior and is people's psychophysiological reactions to internal or external stimuli" (Zhang et al. 2020, p.104).

Emotions are reflected in an individual's mental state and they arise spontaneously rather than through conscious effort (Shu et al. 2018). They are triggered by external stimuli, which the individual perceives as 'personally significant' (Tyng et al. 2017). Researcher often distinguish between basic or primary emotions and complex emotions composed of basic emotions (Russell 1980). Although not undisputed, a dominant concept of emotions states that six basic emotions exist consistently across cultures: joy, love, sadness, fear, anger, and surprise (Plutchik 2001; Russell 1980; Shaver et al. 1987). However, Shu et al. (2018) discuss several other models that have gained substantial consideration in the research community, which include many more basic emotions and also mixed facets of these. Even though there is no unanimously agreed on model of emotions, it remains undisputed that human emotions are an important key to understanding human behavior (see e.g., Elster 2009). Thus, research on automatically recognizing human emotions precisely and quickly will remain the target of considerable effort in scientific research and industry (Shu et al. 2018), as emotions play a vital role in people's decision-making, perception and communication (Zhang et al. 2020).

## 2.2 Recognizing Emotions Using Artificial Intelligence

Although the concept of emotions remains debated, some aspects of it have been generally accepted. There is unanimous agreement that emotions result in physiological reactions such as an increase in heart rate, sweating, and dilation or constriction of blood vessels (i.e., blushing and paling) (Soleymani et al. 2015; Zhang et al. 2020). Many of these physical reactions are measurable, such as physiological changes in human organs and tissues such as heartbeat, skin reactions, blood flow, muscle tensions, facial expressions, changes in voice, etc. (Shu et al. 2018). The ability to (automatically) measure reactions which a human being can hardly control opens many doors for research, especially using AI methods. Specifically, if an algorithm can identify the emotional status of a human counterpart accurately, it would have a wide array for applications in the service industry.

Research on basic emotions has been carried out for some time and it has been shown that these are detectable utilizing one or more visual or auditory modalities and applying machine learning techniques (Zhang et al. 2020). A dominant approach in AI to detect emotions is the analysis of facial expressions (Adolphs 2002) and lately specifically micro expressions (Matsumoto and Hwang 2018). Tivatansakul et al. (2014) for instance, achieve an accuracy of about 86% in detecting the six basic emotions based on facial expressions. Electroencephalogram (EEG) signals also provide meaningful indications concerning the emotional conditions of individuals (Proverbio et al. 2013). Zhang et al. (2020) reported that EEG electrodes applied to the frontal lobe of a subject revealed a classification accuracy of more than 90% across different emotional states. However, using EEG headgear for signal measurement is inconvenient for the individual and it generally takes considerable time to measure the multichannel signals. Consequently, user acceptance is a problem for EEG-based emotion recognition and cannot be used in a real-life service interaction.

Another approach is to identify emotions via speech analysis as suggested by Scherer (1995). Early work in this area by Amir and Ron (1998) provided an automatic classification algorithm to identify emotions in speech. Their effort was limited by the technological capabilities at that time. As technology improved, numerous approaches to emotion recognition through speech have emerged since then. Some years later, Schuller et al. (2003) were already able to distinguish seven emotional states with an accuracy of about 78% using speech analysis. This indicates that with increasing computational abilities, analysis of the human voice is a promising route to detect the emotional status of the speaker.

## 2.3 Emotion Detection Through Speech Analysis

Where emotion recognition is conducted on speech, research often utilizes acoustic features such as amplitude, pitch and formants (Demircan and Kahramanlı 2014; Slimi et al. 2020). A prevalent methodological approach involves the extraction of spectral features, namely the Mel-frequency cepstral coefficients (MFCCs). First introduced by Davis and Mermelstein (1980), MFCCs proved to be particularly promising and computationally efficient for recognizing patterns in speech (Dave 2013; Fraser et al. 2016; Gupta et al. 2018; Logan 2000). Additionally, MFCCs are increasingly used in speech emotion recognition (SER) as the basis for analyzing an individual's emotions (Kishore and Satish 2013; Lalitha et al. 2015; Pakyurek et al. 2020).

Hidden Markov models (HMM) have been used for speech emotion recognition for more than two decades. Utilizing this method, Schuller et al. (2003) identified seven emotional states with an accuracy of about 78%. Aiming for higher recognition accuracy more recent approaches apply deep learning (DL) methods. Examples are recurrent neural networks (RNNs) (Trigeorgis et al. 2016) and combinations of convolutional neural networks (CNNs) and RNNs (Soleymani et al. 2015; Zhao et al. 2019). Zhao et al. (2019) achieve an average test accuracy of around 96% for seven emotional states based on log-Mel spectrograms.

Besides spectrogram features like MFCCs, or pitch, end-to-end architectures use contextual features (Fayek et al. 2017; Trigeorgis et al. 2016) for increased accuracy performances. Fayek et al. (2017) distinguish five emotional states with an accuracy of around 65% through this approach.

While the SER systems focus on a single person, Majumder et al. (2019) analyzes the conversations of a set of individuals and uses this data for emotion classification. Their approach can classify six emotions with an average accuracy of 64%.

This brief review of important studies which used speech analysis to recognize emotions underscores that this is an upcoming and promising area of research. Technological progress, especially increasing computational performance will enable further research.

## 2.4 Automated Detection of Persuasion, Deception and Lies

The aforementioned researches are targeting different sets of basic emotions. When it comes to the detection of truthfulness, research becomes scarcer. It would be of great benefit to the service industry to be able to tell whether a customer is making a point s/he considers truthful or not. As Ekman (2004) points out, lying is a construct of several emotions, dominantly fear, guilt and delight. Therefore, if AI can detect emotions, it must also be possible to detect whether a person is speaking the truth.

In a large scale experiment, using only humans for detection, Bond Jr and DePaulo (2006) found that untrained people were able to correctly identify deception with an average rate of 54%. Their accuracy was higher for audio only than for video and audio information. Similar results are presented by Levitan et al. (2020) wo developed a game, where the participants had to judge whether another person speaks the truth, solely based on audio files. The population of participating laypersons achieved a detection accuracy of 49.93%, i.e., merely chance. Therefore, the aim for an automated system must be to achieve a higher and sustainably more accurate detection rate.

Ortiz (2010) worked on *persuasion* detection from conversations using a data set of transcripts taken from conversations in various hostage situations in the United States. The author concludes that although ML techniques can be utilized to identify the persuasive power from speech, the technical abilities available at the time were only able to train poor classifiers. Park et al. (2014) aim to recognize a person's conviction using visual, acoustic, and verbal descriptors. The authors created dataset of 1,000 film review videos, classified in different descriptor combinations. They observe that especially MFCC features representing low-frequency regions help predicting the actual persuasion. Nojavanasghari et al. (2016) were among the first to apply a deep neural network (DNN) to classify persuasion. They conducted experiments in various combinations involving visual, auditory and textual modalities, on an architecture consisting of several fully connected layers. The authors obtain an accuracy of around 90% utilizing all three modalities combined in a late fusion process.

Bhamare et al. (2020) also used a deep neural network and by utilizing a combination of 20 MFCC features and 68 facial microexpressions were able to detect *deception* with an accuracy of almost 80%. By combining features from video, audio, and text along with Micro-Expression features, Krishnamurthy et al. (2018) detect deception in real life videos with an accuracy of 96.14% tested on a publicly available dataset of 121 video clips from courtroom trials. Using the same dataset, Mathur and Matarić (2020) applied unimodal support vector machines (SVM) and SVM-based multimodal fusion methods to identify deceptions in real-life situations and achieved ~91%. accuracy.

Focusing on the automatic detection of right-out *lies*, Nasri et al. (2016) accomplished an accuracy of almost 80% in detecting a lie. They used 13 MFCC features employing a support vector machine. The data analyzed consisted of both true and untrue narratives of 80 to 120 seconds expressed by 40 volunteers. Karpova et al. (2020) conducted an experiment where 93 volunteers who purposefully lied and tried to hide this from a polytrophic test. Using video analysis and eye-tracking techniques, the authors trained an end-to-end convolutional neural network. The best model achieved a mean balanced accuracy of 64% to detect truth and lie.

Speech-based emotion analyses using MFCCs has great potential to serve as lie detectors (Chamoli et al. 2017). Deep neural network architectures incorporating MFCC features in their classifications have already been able to demonstrate promising results in this regard (Bhamare et al. 2020).

## 2.5   Research Objective

As discussed above, AI methods have been applied to detect emotions from various human expressions, including speech-analysis, with good prediction accuracy. However, published research on detecting whether what a person says represents her/his convictions is scarce and have not yet delivered the envisioned results. To close this research gap in the field of speech-based persuasion detection, the aim of this study is to develop an adequate AI model that detects whether a speaker is convinced that his/her statements on a certain topic are truthful.

For this purpose, supervised learning methods are employed, more precisely, we train models based on a CNN-LSTM hybrid, which recognizes patterns in speech based on injected MFCC features. In this paper, we solely focus on such features, as they provide an efficient and rapid computation tool for recognizing patterns from audio files while being very robust against poor recording conditions (Dave 2013). Moreover, Nojavanasghari et al. (2016) obtained satisfactory results in the persuasion detection domain, including 24 MFCCs in their approach. For our recognition task, we train eight models, each undergoing a three-fold-cross validation, first employing 13 MFCC features, and another eight models utilizing 40 MFCC features. The objective is to determine for what quantity of MFCC features and model hyper-parameters the most accurate results are achievable.

# 3 Research Method

Scientific research aiming to produce an artefact, such as the AI model in focus in this paper, must adhere to a sound methodological framework. Design science research (DSR) (Peffers et al. 2007) is a design-oriented research methodology to solve a specific problem. DSR is widely employed in information systems and computer science demonstrated the capabilities of the complete architecture of the artefact and discussed potential practical applications and further research opportunities. This serves as proof-of-concept that acoustic analysis and specifically spectral features incorporating AI methodologies can be used to detect whether a speaker is talking to her/his true conviction.

As the AI model presented in this paper is not yet in productive use in practice, *evaluation* is only possible by comparing it with other published models. To complete the DSR research, as well as in service design and service research (see e.g. Sudbury-Riley et al. 2020; Teixeira et al. 2019). The methodology is used to properly design and evaluate a value-adding artefact. As DSR deals with constantly evolving technologies and increasingly complex contexts, the approach is particularly well suited to the development of new types of service solutions (Teixeira et al. 2019). DSR is a process of six major steps as illustrated in Figure 1.

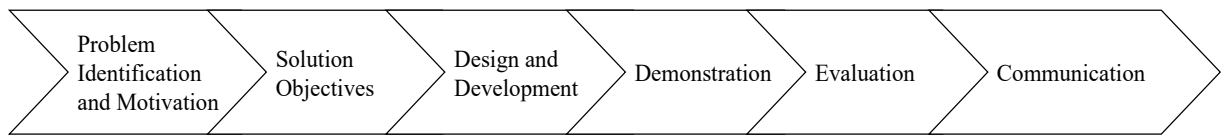

Figure 1: DSR process as suggested by Peffers et al. (2007)

In this research paper, the first step of the DSR process, *problem identification and motivation,* is provided in the introduction. The second step, the definition of the *solution objective,* is provided in the discussion of related work and the description of the research gap. The *design and development* phase starts with an explanation of the setting of the experiment to gather the required data and continues into the actual development process for the suggested artefact, the AI model, through all its stages. In this phase, the overall system architecture, the required pre-processing and the subsequent feature extraction is explained.

To *demonstrate* the functionality and usefulness of the developed AI model, we held workshops with practitioners from the service industry and academic experts on AI and automatic speech recognition. In these workshops, we process, our findings will be *communicated* in scientific and practitioner-oriented journal articles, as well as in presentations and workshops for the scientific and practitioner community to ensure theory-to-practice knowledge transfer.

# 4 Data Collection

Publicly available data set in German language are scarce (Xu et al. 2020). Therefore, to achieve our stated research purposes, we had to generate our own data. We collected an audio dataset consisting of 40 different verbal representations of opinions on four different polarizing topics by 40 different speakers.

## 4.1 Data Collection Process

In order to generate audio data by individual speakers who were clearly identifiable as arguing for or against their conviction, we chose the setting of a debating society / debate club. We distributed recruitment flyers in local schools and universities and on various social media channels inviting volunteers 16 years of age or older to participate in our study, which we referred to simply as an experiment, offering each participant a 10 Euro voucher for Amazon.

The experiment followed typical debate club procedure. The participants were randomly grouped into pairs. Each pair received a random assignment to one of the four polarizing topics (see 4.3 below), and whether they should argue pro or contra.

The participants had 30 minutes to prepare and were instructed not to use the internet in their preparation. Participants were permitted to make bullet point notes but no fully written arguments. Each participant was then asked to argue their assigned position (pro or contra) on their topic in approximately 2 to 3 minutes using free speech. We did not provide the participants with any information about the study's objective beforehand to avoid possible bias. Notably, none of the participants asked about the underlying intention of the experiment.

After the debate, the research team gave each participant feedback on their rhetoric performance / argumentation to help them improve their soft skills. Finally, participants were asked to provide suggestions on improving the operational handling of the experiment. No further interaction with the research team occurred to reduce social desirability response bias for the following step: Immediately after the feedback discussion, participants received an automated email, asking them to complete a brief online survey to determine their basic demographic data and their true conviction about the topic they debated, which they were encouraged to complete as soon as possible. All participants fully completed the survey.

The experiment took place in Germany from December 2020 to January 2021. Due to the prevailing Covid-19 situation, the experiment was conducted online via videoconference. For data privacy purposes, only the audio stream was recorded and used for the research, the video stream was deleted. All participants were informed about the data privacy precautions and signed a corresponding consent form before the experiment started.

## 4.2 Demographics of the Participants

The participants were predominantly high school and undergraduate university students. All debates were conducted in German. 18 women and 20 men between 16 and 59 years (average 23.8 years) participated, including two university lecturers aged 50 and 59, who had no further involvement in this study.

We conducted 20 debates, thus collecting 40 audio recordings. Due to extremely low audio quality (breaks ups, transmission delays and interference due to bad internet connection), two recordings had to be removed, leaving us with 38 individual audio files. Overall, 20 pro and 18 contra debate statements were included in our study.

## 4.3 Topics

We deliberately chose easy and highly polarizing topics for the discussions. The topics (translated to English) were stated as follows:

- **Topic 1**: Should the death penalty and public executions be reintroduced in Germany?
- **Topic 2**: Should cost-covering tuition fees be charged in Germany? (Note: this means approx. 8,000 to 15,000 EUR per semester)[1].
- **Topic 3:** Should the use of hard drugs such as heroin and crystal meth be legalized in Germany?
- **Topic 4:** Should restaurant chains serving unhealthy fast food, such McDonald's or Burger King, be banned in Germany?

We selected these topics in order to increase the likelihood that participants would have a clear and unambiguous opinion. Analysis of the participants' responses to the email questionnaire revealed that the speakers' true convictions were as the research team expected them, i.e., none of the participants supported reintroducing the death penalty etc. However, it is always possible that participants do not state their true position due to social desirability response bias. We address this limitation of our research in Section 7.

The distribution of the topics and the respective pro and contra positions to the participants is provided in Table 1. Table 2 shows how many participants (per topic) argued for or against their conviction.

Of the 38 usable audio files collected, 18 were of participants arguing in accordance with their conviction and 20 where of participants arguing against their conviction. Of the twenty pro debate statements, five participants argued for their conviction (two women and three men) and fifteen participants argued against their conviction (eight women and seven men). Of the eighteen contra debate statement, thirteen participants argued for their conviction (seven women and six men) and five argued against their conviction (two women and three men).

# 5 Development of the AI Models

## 5.1 Pre-Processing and Feature Extraction

In order to train a classification model, entries must be converted into an analyzable format and adequate metrics for speech analysis must be identified and extracted. This can be achieved in two ways: One approach, known as automatic speech recognition (ASR), involves analyzing speech at the acoustic level by directly interpreting occurring speech characteristics based on the audio file. The other approach, known as natural language processing (NLP), involves analyzing speech at the linguistic level by transcribing the audio file into text format and subsequently identifying and analyzing its features on the syntactic, lexical and semantic (content) level.

|  | Topic 1 | Topic 2 | Topic 3 | Topic 4 | Sum |
|---|---|---|---|---|---|
| **Pro-position** | 5 | 6 | 5 | 4 | 20 |
| **Contra-position** | 3 | 4 | 5 | 6 | 18 |

Table 1: Allocation of topics and assigned position

---

[1] Please note that education is free in Germany and German public universities do not charge tuition fees.

| | Topic 1 | Topic 2 | Topic 3 | Topic 4 | Sum |
|---|---|---|---|---|---|
| **Argumentation reflected the speaker's conviction** | 4 | 5 | 2 | 7 | 18 |
| **Argument did not reflect the speaker's conviction** | 4 | 5 | 8 | 3 | 20 |

Table 2: True conviction of the participants per topic

In this study, we adopt the first approach: the analysis of acoustic speech features. As discussed in Section 2.3 above, we extracted spectral features called Mel-frequency cepstral coefficients (MFCCs) (Davis and Mermelstein (1980). Using MFCC processing has two advantages: (1) it provides computational efficiency for recognizing patterns in speech (Dave 2013; Fraser et al. 2016; Gupta et al. 2018; Logan 2000), and (2) it generates strong results even using audio files of inferior quality (Tivatansakul et al. 2014). The second advantage is very important in our study because the recordings of our data set were made via videoconferencing with commercial video cam microphones rather than in a laboratory environment using high end microphones, as originally planned pre-COVID19. From a practical standpoint, the mediocre quality of the audio files is a realistic approximation of recorded or live audio in real-life environments, such as in the service industry.

### 5.1.1 Data Augmentation

Due to the limited number of datasets, we augmented the dataset through automated data manipulation, using time stretching, pitch shifting, and the addition of noise to generate 304 speech files based on the initial 38 datasets, which we pre-processed as described below.

### 5.1.2 Data Pre-Processing

In a pre-processing stage, we transformed the speech signal from analogue waveform of the spoken language into feature vectors. First, we split the continuous speech signal into discrete frames or windows of equal length. Based on the fact that speech is considered static over a period of 5 to 25 milliseconds, a window size of 20 milliseconds is commonly used (Logan 2000). This results in a fixed range that can be analyzed, also known as short-time spectral analysis. The audio files we collected were on average 144 seconds long. Second, we analyzed frames with the default hop size of 512, which corresponds to 20-millisecond window size, which provided unsatisfactory results. We then compared the results using larger window sizes, generating the best results with a hop size of 8192, which corresponds to a window size of 186 milliseconds. Our hypothesis is that a 20-

millisecond window size provides less accurate patterns of a speaker's true conviction than a significantly larger window size, such as 186 milliseconds. This hypothesis requires more thorough testing. Third, since individual windows usually overlap, resulting in signal jumps at the edges and erroneous results in the subsequent frequency analysis (Tiwari 2010), we removed the edge effects using the Hamming-Function (see Logan 2000).

### 5.1.3 Feature Extraction Using MFCCs

After pre-processing the data, we extracted and analyzed spectral features called Mel-frequency cepstral coefficients (MFCCs) (Dave 2013; Dessouky et al. 2014) from the audio data. Since their introduction by Davis and Mermelstein (1980) over 40 years ago, MFCCs are still considered state-of-the-art in speech recognition research (Fraser et al. 2016; Gupta et al. 2018). MFCC extraction and analysis represents the speech amplitude spectrum concisely, and provides highly robust and accurate results even under unfavorable recording conditions (Dave 2013; Logan 2000).

Researchers also favor MFCCs for speech feature analysis because they rely on the Mel-scale frequency range, which reflects the frequency range of human hearing (Verde et al. 2018). According to the definition of the 'Mel' unit, two sounds perceived as equally distant from each other have the same tonality value. The human ear does not perceive all frequencies equally: Frequencies above approximately 1,000 Hertz are typically less noticeable.

Accordingly, below 1,000 Hertz the ratio between frequency and Mel scale is linear and above 1,000 Hertz the ration is logarithmic. This characterizes Mel as a unit of perceived pitch (Stevens et al. 1937).

Human utterances are composed of excitation signals generated through the glottis, shaped by an impulse response from downstream filters in the vocal tract. The shape of the sound defines the sound (Finch 2016). When this shape is accurately identified by an ASR system, a precise representation of the individually produced phoneme becomes possible and thus analyzable (Logan 2000). A phoneme denotes the smallest entity conveying a segregated meaning within a language's phonetic system. Phonemes are thus phonological objects, whereas undirected individual speech sounds are referred to as phones, which are the smallest segmental phonetic units of speech (Finch 2016).

In order to determine the shape of an audio signal using spectral analysis, the time-domain-based audio signal must first be transformed into the frequency domain. This is achieved by applying the fast fourier transform (FFT) algorithm (Kishore and Satish 2013). This results in a representation of the power spectrum of each frame, allowing the frequencies within the frames to be identified more precisely than by analyzing the time-domain-based signal directly (Logan 2000).

To account for the varying sensitivity in the feature extraction, the results of the FFT algorithm are then mapped to the Mel scale (Ittichaichareon et al. 2012) using the Mel filter bank, which is a set of filters. Each filter collects the energy of one frequency range; the spectral components are thus divided into different frequency groups. Subsequently, the logarithm of these collected Mel values is formed. This approach also has its origin in human hearing since the human ear also reacts logarithmically to different signals; hence, differences at high frequencies are perceived less arbitrarily than differences at low frequencies. By using the logarithmic values, a more robust recognition is guaranteed, which in addition is less vulnerable to variations of the linguistic input data (Dave 2013).

To extract features accurately, it is crucial to have a method of separating source and filter in order to be able to only reuse the filter information. This is achieved by applying the discrete cosine transformation (DCT) to the Mel filter bank outputs (Ranjan 2019). This transformation classifies the coefficients according to their significance (Dave 2013).

A set of MFCCs is thereby calculated for each speech window. Generally, the first thirteen MFCCs, which represent the envelope of spectra, are considered in an analysis. Twelve of these parameters are related to the amplitude of the frequencies. The total energy of each frame is added in a thirteenth feature, since it correlates with the spoken phoneme, thus providing additional support for recognizing different phonemes (Dave 2013).

Speech signal are dynamic and subject to constant change. However, since the cepstral coefficients only contain the information of a certain frame, this dynamic cannot yet be mapped. The information about the temporal dynamics of the signal is acquired by calculating the first- and second-order derivatives of these coefficients for each of the thirteen cepstral coefficients. These derivatives represent the change in cepstral values within two adjacent frames. Consequently, an MFCC vector consists of 40 features (Yu and Deng 2016).

To determine whether and to what extent the number of MFCC features is reflected in the performance of the models trained, first a feature vector consisting of 13 and then a feature vector consisting of 40 MFCC features per recorded dataset is extracted. As the debates vary in duration, each vector is zero-centered and scaled between -1 and 1 during the extraction process. This finalizes the pre-processing of the data and enables the AI models to be trained with the respective MFCC feature vectors.

## 5.2 Training of the AI Models

The proposed architecture follows a hybrid approach, combining a convolutional neural network (CNN) and a long short-term memory (LSTM) neural network, hereafter referred to as CNN-LSTM. CNNs have been chosen over deep neural networks (DNNs) because they explicitly exploit the spectral feature space's structural locality. CNNs use joint weighting filters and pooling to provide improved spectral and temporal invariance properties to the model. As a result, they typically produce better-generalized and more robust models than DNNs (Zhang et al. 2017). The LSTM was included due to its high performance in handling time-series correlations (Sainath et al. 2015) which aided in overcoming the time-series problem.

A general overview of the proposed network architecture for the voice-based recognition of whether what a speaker says represents his/her convictions is provided in Figure 2. Here the respectively generated MFCC feature vectors provide the baseline in detecting a speaker's conviction. Next, these pre-processed MFCC feature vectors are fed into a time-distributed

1D convolutional layer. Subsequently, the CNN output is handed over for further processing within the following LSTM layer. In the last layer, all neurons are fully connected to each other. Here the final determination of whether the speaking person is convinced that the statements he/she is making are true or not, assigned to two discreet options, zero or one.

By randomly shuffling the data based on their indices, a predominance of augmented and non-augmented data in both training and test sets is ensured. These randomizations are performed simultaneously three times for each feature vector. This permits three-fold cross-validation, which is later applied to each of the models utilizing 13 MFCC and the models incorporating 40 MFCC features. Subsequently, the mean value of the performance metrics employed for all trained models is computed across those three folds in order to validate functionality. For each fold, the data set is split into 70% training data and 30% validation data.

Given the time series problem, the input dimension had to be adapted so that the time variable, i.e., the segments, are loaded into the model first, followed by the number of MFCCs to create an appropriate model.

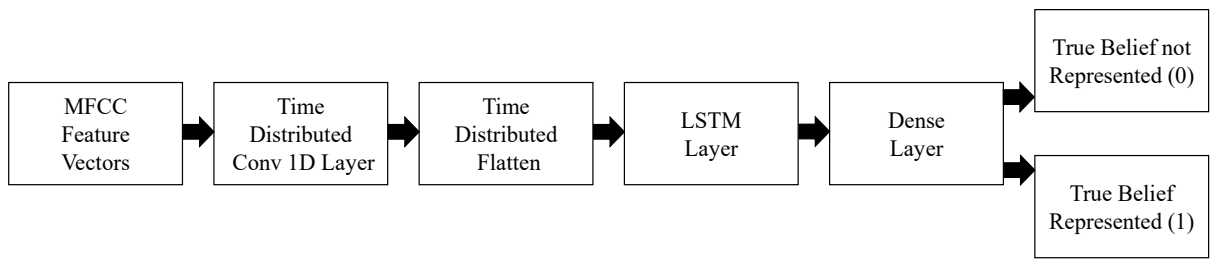

Figure 2: Proposed CNN-LSTM architecture

As illustrated in Figure 2, the proposed model proceeds sequentially. The inclusion of a time distributed convolutional layer requires that every input is adapted to this layer. This procedure serves to identify movements and directions with chronologically ordered data, such as in the scenario presented in this paper. Consequently, the Time Distributed Layer provides the possibility to feed sequences as inputs and therefore allows to include individual sequences in the predictions (Keras 2021).

The rectified linear unit (ReLU) is employed as the activation function in all scenarios. The padding has been set to the value 'same', which means that by adding zeros at the borders of each feature vector, the output will match the input size of each feature vector from the convolutional layer. Experiments were performed with combinations of 16 and 32 filters and kernel sizes of 5 and 20.

Subsequently, the CNN output is flattened to reduce the dimension, thus allowing the LSTM to accept it for further processing. For LSTMs, it is common to use tangent hyperbolicus (tanh) as an activation function to push the values between -1 and 1 (Saxe 2018). The architecture proposed also utilizes this activation function regarding the LSTM. Furthermore, we built models with either 20 or 40 input neurons for the LSTM. The output layer is fully connected and utilizes, as is customary, the logistic function as the activation function (Saxe 2018). The number of predominant labels are specified here as either zero, indicating that the subject does not truly believe the statement being expressed, or one, denoting that the statement does in fact correspond to the subject's true conviction. (Kingma and Ba 2015). The optimizer employed is ADAM in its standard configurations.

Upon initial analysis, the number of epochs was set at 250 and the batch size was set at 16 in this study. Since experiments were conducted applying different amounts of extracted MFCCs, varying numbers of filters, diverse kernel sizes, and multiple amounts of LSTM input neurons, it was possible to train both 8 models containing 13 MFCCs and 8 models containing 40 MFCCs. This process involved subjecting each model to three-fold cross-validation, resulting in 48 models. The three-fold cross-validation was carried out to assess the robustness of each model. The performance metrics displayed in Table 3 and Table 4 show the obtained results for each model over the three runs.

# 6 Results

This section outlines the chosen performance metrics and presents the average performance of the trained models. The metrics are applied to models using 13 MFCC features, followed by those incorporating 40 MFCC features. Thus, the results regarding all 16 neural network architectures differing in the number of filters, kernel size, and LSTM input neurons are presented. This emphasizes and highlights the final model's importance as part of a dynamic process of model comparison.

## 6.1 Performance Metric Parameters

The performance metrics are applied to measure each model's efficiency and reliability separately. Four key parameters form the basis of the performance assessment:

**True Positive** (TP): The number cases in which the model correctly categorize the conviction of a speaker as represented by the position he/she took in the debate.

**False Negative** (FN): The number cases in which the model did not correctly categorize the conviction of a speaker as represented by the position he/she took in the debate.

**True Negative** (TN): The number cases in which the model correctly categorized the conviction of a speaker as not represented by the position he/she took in the debate.

**False Positive** (FP): The number cases in which the model did not correctly categorize the conviction of a speaker as not represented by the position he/she took in the debate.

These parameters alone are not sufficient to fully validate the results. For example, a high TP rate can be achieved when there is an extreme imbalance in the ratio between recordings of people whose convictions were not represented and speech data of people whose convictions were represented. For example, a ratio of 1:10 could lead to a very high TP rate even if the learning model performs poorly overall. To avoid this issue, a balanced ratio of input data is required, as well as a weighted consideration of the TP and TN rates (Aldwairi 2018).

In order to assess the models' performance holistically, further parameters were analyzed, namely accuracy, precision, sensitivity, and, $F_1$-score (Saxe 2018). **Accuracy** reflects the correctly classified proportion of a model composed of all data points. It is calculated by dividing the count of TP and TN by the count of TP, TN, FP, and FN. **Precision** reflects the share of TP-classified samples in the total number of samples classified as positive (TP + FP). **Sensitivity**, also referred to as true-positive rate, demonstrates a model's predictive power in terms of the proportion of correctly identified TPs to the total number of true-positive observations (TPs + FNs). The **$F_1$-score** complements the accuracy measures because it balances unequal class distributions within a training dataset by calculating the harmonic mean between precision and sensitivity measures.

## 6.2 Set 1: 13 MFCC Features

The metrics reported in Table 3 were derived through three-fold cross validation. All models were trained on 13 MFCC features but with varying hyper-parameters (filters, kernel size and LSTM neurons). The best model trained, M1-13, achieved an accuracy of 72.46% that we use a multi-minute utterance to detect whether what a speaker is saying represents her/his conviction. To achieve this, we analyzed 186 millisecond periods, which is much longer periods than the 20 milliseconds typical used in automatic speech recognition or emotion detection.

## 6.3 Set 2: 40 MFCC Features

To advance the models prediction accuracy, we included additional MFCC features in the training. As illustrated in Table 4, increasing the number to 40 MFCC characteristics significantly improved the prediction power across all applied performance metrics. The hyper-parameters involved were adjusted to match those for the models with 13 MFCCs, ensuring that all models were subjected to a threefold cross-validation. The best trained model, M8-40, achieves a satisfactory accuracy of 98.12%.

| Model | Filters | Kernel Size | LSTM Neurons | Ø Accuracy | Ø Precision | Ø Sensitivity | Ø $F_1$-Score |
|-------|---------|-------------|--------------|------------|-------------|---------------|---------------|
| M1-13 | 16 | 5 | 20 | 72.46% | 78.99% | 62.13% | 68.89% |
| M2-13 | 32 | 5 | 20 | 57.61% | 61.90% | 81.44% | 66.62% |
| M3-13 | 16 | 20 | 20 | 58.33% | 57.94% | 90.09% | 68.85% |
| M4-13 | 32 | 20 | 20 | 50.36% | 52.69% | 83.61% | 64.34% |
| M5-13 | 16 | 5 | 40 | 52.17% | 56.65% | 75.24% | 60.01% |
| M6-13 | 32 | 5 | 40 | 63.41% | 65.51% | 60.54% | 60.87% |
| M7-13 | 16 | 20 | 40 | 61.23% | 62.38% | 81.16% | 67.84% |
| M8-13 | 32 | 20 | 40 | 64.49% | 58.84% | 68.11% | 61.10% |

Table 3: Scores for all model using 13 MFCC features

averaged over all three folds.

The precision and $F_1$-score of this model confirm it to be the best of all tested models utilizing 13 MFCCs. With a sensitivity of 62.13%, the model predicts about 62 of 100 speeches correctly as positive (TP) and 28 of 100 speeches incorrectly as positive (FP).

However, the results are not satisfactory. This was surprising because the extraction of the first 13 MFCC features is a well-established method in speech pattern detection (Park et al. 2014). One of the main differences of our approach is,

Thus, by including 27 additional MFCC features, we were able to increase the accuracy of our model by over 34%. Additionally, this model outperforms all other trained models by achieving a precision of 100%, meaning this model correctly classified all TPs from the test data provided. Moreover, this model has a sensitivity level of 98.15% and an average $F_1$-score of 99.05% across all folds which demonstrates that model M8-40 is the most accurate model in our research.

| Model | Filters | Kernel Size | LSTM Neurons | Ø Accuracy | Ø Precision | Ø Sensitivity | Ø F$_1$-Score |
|-------|---------|-------------|--------------|-----------|-------------|---------------|---------------|
| M1-40 | 16 | 5 | 20 | 80.80% | 65.94% | 73.88% | 67.66% |
| M2-40 | 32 | 5 | 20 | 74.78% | 67.61% | 97.16% | 82.58% |
| M3-40 | 16 | 20 | 20 | 78.64% | 84.47% | 90.81% | 84.80% |
| M4-40 | 32 | 20 | 20 | 91.67% | 77.07% | 96.73% | 84.32% |
| M5-40 | 16 | 5 | 40 | 90.94% | 84.97% | 94.08% | 88.48% |
| M6-40 | 32 | 5 | 40 | 84.06% | 57.07% | 64.67% | 60.34% |
| M7-40 | 16 | 20 | 40 | 80.80% | 85.90% | 77.01% | 80.65% |
| M8-40 | 32 | 20 | 40 | 98.91% | 100.00% | 98.15% | 99.05% |

Table 4: Scores for all model using 40 MFCC features

All models trained on 40 MFCC features perform significantly better than those trained on 13 MFCC characteristics. On average, these models achieved 85.08% accuracy, leading to a 25% improvement compared to the models trained with the 13 MFCCs traditionally used in ASR.

# 7 Evaluation of the Proposed AI Model

The proposed CNN-LSTM architecture addresses the automatic detection of whether or not what a speaker is saying represents her/his convictions based on spectral features, the MFCCs. Due to the absence of directly comparable published research, the critical evaluation using a vis-à-vis comparison is not possible.

==[A note: The authors keep monitoring publications in this research area and will update this section of the manuscript if another comparable research will be published.]==

Recent research on lie detection conducted by Nasri et al. (2016) most closely resembles our classification approach. Incorporating MFCC features, the authors perform a binary classification utilizing an SVM, achieving nearly 80% accuracy for lie detection. The accuracy achieved in their approach is outperformed by the model proposed in this research when incorporating 40 MFCC features in the classification. As Nasri et al. (2016) did not provide information on the required resource consumption, it is not possible to evaluate whether our approach performs with lower performance requirements.

# 8 Discussion: Application of Findings in the Service Industry

Customers' emotions play a crucial role in the service industry (Mattila and Enz 2002). The behavior of frontline service personnel is a vital component of customers' evaluation of the service received (Hartline et al. 2000). Especially in an environment where there is no physical interaction between customer and service agent, like in a call center, it is critical that the agent is aware of the customer's emotional state, in order to act accordingly.

Many researchers have investigated methods of automatically detecting human emotions, mainly through face recognition, EEG etc. The analysis of the customers' voice is a comparatively young field of research. Researchers have had good success detecting basic human emotions automatically via speech analysis, but there are few findings reported on successful automatic detection of complex emotions (i.e., combinations of several basic emotions).

A complex emotion of interest to the service industry is 'lying'. According to Ekman (2004), lying is a complex emotional construct that is interdependent with other emotions and itself triggers emotions in the liar and in the person being lied to (Proverbio et al. 2013). Not telling the truth comes in many different flavors: There is the straight lie, but also deception, bluffing, swindle, or the white lie, and many more. They all have in common that the customer is telling the agent a story which does not accurately reflect her/his conviction of what is true.

As research into using speech analysis to detect whether a speaker is lying is still in its infancy, our work started with a broader focus: We investigated whether the statement a speaker provides reflects her/his true convictions.

## 8.1 Contribution to Practice

The ability to automatically detect whether a person is speaking to her/his conviction offers numerous applications to the service industry, both on the organizational as well as on the individual level.

### 8.1.1 Organizational Level Applications

Untruthful statements pervade organizational life and represent significant challenges in negotiations, job interviews, expense reporting, corporate accountability, and so on. For example, the German Insurance Association estimates that every tenth insurance claim has a fraudulent component and reports that around 5% of all customers have previously submitted a fraudulent claim. The damage caused by fraudulent claims to German insurers is estimated to exceed 5 bn Euro in 2020 (GDV 2020). While no industry-wide figures are available about fraudulent travels claims and expense reports, industry experts estimate that approximately 15% of all German companies are affected by fraudulent claims valuing on average approximately 600 EUR per employee per year (Romberg 2019).

Applying AI techniques to detect whether a caller is speaking to her/his conviction when making a claim may not completely solve these issues, but it could help flag the caller for more intensive questioning and give her/him the opportunity to correct his/her statements and avoid submitting a fraudulent claim.

Organizations holding job interviews can also benefit from such an application. Apparently, it is common practice these days to bend the truth in job interviews (Kelly 2021). As initial interviews of suitable candidates are often held via telephone, automated analysis of these verbal interactions can enable the organization to investigate if the candidate is stating things s/he believes are truthful, saving both sides time, effort and disappointment. Widespread use of such mechanisms may even motivate candidates to return to the good virtue of only giving true statements in job interviews.

Automated emotion recognition may also be useful in business negotiation situations, where lying and bluffing are a common part of the process (Kaufmann et al. 2018). If an algorithm is able to tell whether what the opposing side is saying represents their convictions, this would strengthen one's own position. This application is certainly difficult, however, because experienced negotiators are highly skilled in hiding their emotions and because using such an application is likely only legal if both parties are informed and consent.

Such an algorithm would also be useful in inhouse sales training. A good salesperson truly believes in the product s/he is selling to the customer. If the person is able to convince the AI algorithm using arguments that are not detected as going against her/his convictions, the person is well positioned to face the external customer.

Social desirability response bias (SDR), i.e., the tendency of informants to answer questions in a manner that will be viewed favorably by others, is a significant challenge in market research and in social science. SDR leads to faulty data in the dataset because the responses do not reflect the person's actual opinion. A prominent example of this effect are the polls for the 2016 presidential election in the USA. One reason for the dramatic difference between poll results and actual voting is assumed to be the so-called "shy Trumpers", suggesting that support for Trump was viewed as socially undesirable, and that his supporters were unwilling to admit their support to pollsters (Mercer et al. 2016). If interviews are conducted via telephone, the algorithm could flag the suspected responses as not trustworthy so they could be excluded from the data set. The resulting 'clean' set would only contain information that the speakers believe to be true.

In the future, further developed versions of this algorithm may offer new ways to verify stated intentions in security-critical areas, such as in check-in / boarding procedures at the airport or other transport facilities. Also, immigration interviews at airports and border passport control (Dando and Ormerod 2020) could potentially benefit, as well as law-enforcement and courts of law (Burzo et al. 2018).

In summary, there are countless areas where automated emotion detection and specifically detection of truthful conviction of statements can be applied on an organizational level, with vast potential to decrease corporate damages and promote honest behavior.

### 8.1.2 Individual Level Applications

The use of automated detection of whether a person is speaking to her/his convictions also has applications at the individual level, specifically in the service industry.

In the service industry, it is an open secret that customers' statements at helpdesks or the customer service desk do not always fully reflect the truth (Alton 2017). It would save time and resources if an algorithm could detect whether the customer rebooted her/his computer before s/he called the helpdesk, or whether the described product malfunction is actually true. Such an algorithm could also help verify whether and when a product was really sent to the support team, resulting in productivity gains and reduction of operational losses.

There are many individual applications in medical and therapeutic fields. Mobile assisted living systems help vulnerable patients manage their daily lives independently. An AI algorithm enquiring how much they drank during the day or whether they took their medication correctly could enhance these systems, because patients often do not to answer these questions truthfully (Fainzang 2002). The algorithm could identify answers that do not reflect the person's convictions and trigger human intervention from a designated caregiver to ensure well-being of the patient. Psychotherapists could employ such an algorithm to see whether the patient is speaking to her/his convictions and physicians could apply the algorithm to improve anamnesis and the resulting diagnosis, especially with regard to socially difficult topics such as sexually transmitted diseases, drug use, or drinking and eating habits (Palmieri and Stern 2009). Naturally, these applications require transparent disclosure to the patient and adherence to corresponding data privacy regulations.

## 8.2 Contribution to Theory

This research empirically demonstrates that AI speech analysis can detect whether what an individual is saying represents her/his true convictions directly from analysis of the speech, without semantic understanding. We developed AI models that demonstrate a remarkable accuracy of up to 98.91% in identifying whether someone is speaking to her/his conviction in an analysis of audio files in a pre-defined debating environment. This research contributes to the body of knowledge by demonstrating the ability to detect conviction automatically by analyzing speech.

This study provides empirical evidence that the defining voice features essential to the speaker's convictions are contained within the verbal statements' spectral representation as measured by the 40 MFCCs we applied. Thus, our findings demonstrate that extending the 13 MFCCs typically used to 40 MFCCs dramatically enhances the outcome quality.

Finally, while extant research demonstrates the use of hybrid CNN-LSTM architectures in basic emotion detection and speech recognition, our study demonstrates that they can also be applied to verify conviction.

# 9 Limitations and Further Research

The most significant limitation of this research is the limited size of the data set, which currently includes only 38 debates. While data augmentation is a solid and commonly accepted approach, a larger number of original audio files from different speakers would be beneficial. Although we chose topics for discussion which are very polarizing to increase the likelihood that participants are for or against them, participants may have hidden their convictions and not reported their true position to us. Finally, our participants were all German speaking and from a German cultural background, potentially limiting the generalizability of our findings to other languages and cultural contexts.

Our findings offer a strong basis for promising further research. First, to confirm our results, similar studies should include a larger sample size, reduce the likelihood of participants providing false statements, and include participants speaking different languages and from different cultural contexts. To enhance the performance and accuracy of our model, we call for research to determine the minimum audio file length needed to reliably verify conviction, and for comprehensive analysis of the accuracy-triggering neurons. Given that 6 out of 8 constructed models containing 40 MFCCs show overfitting at some point in the learning process, further optimization of the number of epochs and the batch size is needed. Further research is also needed to detect patterns in the selection of parameters with respect to the phenomenon of memorization. Moreover, to increase the predictive accuracy, further research should also analyze transcripts of the audio files as described by Zhang et al. (2010) and merge the results with those gained from the purely acoustic analysis to provide an enhanced multi-modality-based and therefore more robust system. This approach could be further enhanced by extracting psychometric dimensions from the transcribed audio files Ahmad et al. (2020). Finally, we recommend that future research considers how individual personality traits such as anxiety and self-confidence influence outcome quality and accuracy of the prediction.

# 10 Conclusion

This paper presents a working model for detecting whether a person is speaking to her/his true conviction using a CNN-LSTM hybrid, exclusively analyzing acoustic spectral features, namely MFCCs. The CNN first reduces the input features' spectral variation and then forwards this to the LSTM layer to perform temporal shaping. We tested the proposed architecture on a specially generated German language dataset consisting of 38 audio files, generated in the experimental setting of a debating club. Out of the 16 models developed over three folds, the best achieved an accuracy and an $F_1$-score of just over 98%. Our results demonstrate the usefulness of applying CNN-LSTM hybrids to perform acoustic speech analysis to detect whether or not a person is speaking to her/his true convictions. While many questions about the underlying patterns in spoken language with regard to the true conviction of a speaker remain, this research indicates that further patterns are recognizable in the deeper MFCC features. The number of runs performed on the individual models and the relatively small dataset prohibit definitive conclusions on this question which underscores the need for further research.

Our findings are applicable to a wide range of processes in the service industry. All spoken interactions, such as on the telephone or in a recorded physical conversation, could potentially be analyzed using the proposed algorithm. Specifically, interactions in which customers are often deceitful, such as insurance claims or job interviews could benefit from an AI that can detect whether or not the speaker is arguing to her/his convictions. Flagging questionable parts of the conversation would enable the agent to follow up more closely on that topic and potentially prevent the customer from making false statements or committing wrongdoing.

Our findings demonstrate that semantic understanding of a conversation is not necessary to detect human emotion. Rather, analyzing the spectral features of the human voice is sufficient. This fact allows the content of the conversation to remain confidential to the human participants, since the AI is not attuned to it. As the old saying goes: "it's not what you say, but how you say it".